\documentclass[aps,prd,superscriptaddress,twocolumn,showpacs,amsmath,amssymb]{revtex4-1}

\usepackage{epsfig}
\usepackage{dcolumn}
\usepackage{bm}
\usepackage{amsmath}    
\usepackage{amssymb}
\usepackage{mciteplus}
\usepackage{subfigure}
\usepackage{graphicx}

\include{include}

\newcommand{\MetRaw}{\mbox{$E\kern-0.60em\raise0.10ex\hbox{/}_{T}^{\rm raw}$}}
\newcommand{\MetVec}{\mbox{$\vec E\kern-0.60em\raise0.10ex\hbox{/}_{T}$}}
\newcommand{\MetVecRaw}{\mbox{$\vec E\kern-0.60em\raise0.10ex\hbox{/}^{\rm raw}_{T}$}}
\newcommand{\MET}{\mbox{$E\kern-0.60em\raise0.10ex\hbox{/}_{{T}}$}}
\newcommand{\METzero}{\mbox{$E\kern-0.60em\raise0.10ex\hbox{/}_{T}^0$}}
\newcommand{\MetSpec}{\mbox{$E\kern-0.60em\raise0.10ex\hbox{/}_{T}^{spec}$}}
\newcommand{\MetRel}{\mbox{$E\kern-0.60em\raise0.10ex\hbox{/}_{T, rel}$}}

\newcommand{\fb}{\text{fb}^{-1}}

\newcommand{\et}{E_{{T}}}
\newcommand{\Et}{E_{{T}}}
\newcommand{\Etzero}{E_{T}^0}

\newcommand{\tofrv}{\text{TOF}^\text{R}}
\newcommand{\tofwv}{\text{TOF}^\text{W}}
\newcommand{\deltat}{\Delta t}
\newcommand{\deltatZero}{\Delta t^{0}}
\newcommand{\deltatRV}{\Delta t^\text{R}}
\newcommand{\deltatWV}{\Delta t^\text{W}}
 
\newcommand{\neutralinoOne}{\tilde{\chi}_1^0}

\newcommand{\gravitino}{\tilde{G}}

\newcommand{\ppbar}{p \overline{p}}

\begin{document}
\bibliographystyle{apsrevM}

 \title{A signature-based search for delayed photons in \\
exclusive photon plus missing transverse energy events from \\ 
$\ppbar$ collisions with $\sqrt{s}=1.96$ TeV}

\date{\today}

\affiliation{Institute of Physics, Academia Sinica, Taipei, Taiwan 11529, Republic of China}
\affiliation{Argonne National Laboratory, Argonne, Illinois 60439, USA}
\affiliation{University of Athens, 157 71 Athens, Greece}
\affiliation{Institut de Fisica d'Altes Energies, ICREA, Universitat Autonoma de Barcelona, E-08193, Bellaterra (Barcelona), Spain}
\affiliation{Baylor University, Waco, Texas 76798, USA}
\affiliation{Istituto Nazionale di Fisica Nucleare Bologna, $^{ee}$University of Bologna, I-40127 Bologna, Italy}
\affiliation{University of California, Davis, Davis, California 95616, USA}
\affiliation{University of California, Los Angeles, Los Angeles, California 90024, USA}
\affiliation{Instituto de Fisica de Cantabria, CSIC-University of Cantabria, 39005 Santander, Spain}
\affiliation{Carnegie Mellon University, Pittsburgh, Pennsylvania 15213, USA}
\affiliation{Enrico Fermi Institute, University of Chicago, Chicago, Illinois 60637, USA}
\affiliation{Comenius University, 842 48 Bratislava, Slovakia; Institute of Experimental Physics, 040 01 Kosice, Slovakia}
\affiliation{Joint Institute for Nuclear Research, RU-141980 Dubna, Russia}
\affiliation{Duke University, Durham, North Carolina 27708, USA}
\affiliation{Fermi National Accelerator Laboratory, Batavia, Illinois 60510, USA}
\affiliation{University of Florida, Gainesville, Florida 32611, USA}
\affiliation{Laboratori Nazionali di Frascati, Istituto Nazionale di Fisica Nucleare, I-00044 Frascati, Italy}
\affiliation{University of Geneva, CH-1211 Geneva 4, Switzerland}
\affiliation{Glasgow University, Glasgow G12 8QQ, United Kingdom}
\affiliation{Harvard University, Cambridge, Massachusetts 02138, USA}
\affiliation{Division of High Energy Physics, Department of Physics, University of Helsinki and Helsinki Institute of Physics, FIN-00014, Helsinki, Finland}
\affiliation{University of Illinois, Urbana, Illinois 61801, USA}
\affiliation{The Johns Hopkins University, Baltimore, Maryland 21218, USA}
\affiliation{Institut f\"{u}r Experimentelle Kernphysik, Karlsruhe Institute of Technology, D-76131 Karlsruhe, Germany}
\affiliation{Center for High Energy Physics: Kyungpook National University, Daegu 702-701, Korea; Seoul National University, Seoul 151-742, Korea; Sungkyunkwan University, Suwon 440-746, Korea; Korea Institute of Science and Technology Information, Daejeon 305-806, Korea; Chonnam National University, Gwangju 500-757, Korea; Chonbuk National University, Jeonju 561-756, Korea; Ewha Womans University, Seoul, 120-750, Korea}
\affiliation{Ernest Orlando Lawrence Berkeley National Laboratory, Berkeley, California 94720, USA}
\affiliation{University of Liverpool, Liverpool L69 7ZE, United Kingdom}
\affiliation{University College London, London WC1E 6BT, United Kingdom}
\affiliation{Centro de Investigaciones Energeticas Medioambientales y Tecnologicas, E-28040 Madrid, Spain}
\affiliation{Massachusetts Institute of Technology, Cambridge, Massachusetts 02139, USA}
\affiliation{Institute of Particle Physics: McGill University, Montr\'{e}al, Qu\'{e}bec H3A~2T8, Canada; Simon Fraser University, Burnaby, British Columbia V5A~1S6, Canada; University of Toronto, Toronto, Ontario M5S~1A7, Canada; and TRIUMF, Vancouver, British Columbia V6T~2A3, Canada}
\affiliation{University of Michigan, Ann Arbor, Michigan 48109, USA}
\affiliation{Michigan State University, East Lansing, Michigan 48824, USA}
\affiliation{Institution for Theoretical and Experimental Physics, ITEP, Moscow 117259, Russia}
\affiliation{University of New Mexico, Albuquerque, New Mexico 87131, USA}
\affiliation{The Ohio State University, Columbus, Ohio 43210, USA}
\affiliation{Okayama University, Okayama 700-8530, Japan}
\affiliation{Osaka City University, Osaka 588, Japan}
\affiliation{University of Oxford, Oxford OX1 3RH, United Kingdom}
\affiliation{Istituto Nazionale di Fisica Nucleare, Sezione di Padova-Trento, $^{ff}$University of Padova, I-35131 Padova, Italy}
\affiliation{University of Pennsylvania, Philadelphia, Pennsylvania 19104, USA}
\affiliation{Istituto Nazionale di Fisica Nucleare Pisa, $^{gg}$University of Pisa, $^{hh}$University of Siena and $^{ii}$Scuola Normale Superiore, I-56127 Pisa, Italy, $^{mm}$INFN Pavia and University of Pavia, I-27100 Pavia, Italy}
\affiliation{University of Pittsburgh, Pittsburgh, Pennsylvania 15260, USA}
\affiliation{Purdue University, West Lafayette, Indiana 47907, USA}
\affiliation{University of Rochester, Rochester, New York 14627, USA}
\affiliation{The Rockefeller University, New York, New York 10065, USA}
\affiliation{Istituto Nazionale di Fisica Nucleare, Sezione di Roma 1, $^{jj}$Sapienza Universit\`{a} di Roma, I-00185 Roma, Italy}
\affiliation{Mitchell Institute for Fundamental Physics and Astronomy, Texas A\&M University, College Station, Texas 77843, USA}
\affiliation{Istituto Nazionale di Fisica Nucleare Trieste/Udine; $^{nn}$University of Trieste, I-34127 Trieste, Italy; $^{kk}$University of Udine, I-33100 Udine, Italy}
\affiliation{University of Tsukuba, Tsukuba, Ibaraki 305, Japan}
\affiliation{Tufts University, Medford, Massachusetts 02155, USA}
\affiliation{University of Virginia, Charlottesville, Virginia 22906, USA}
\affiliation{Waseda University, Tokyo 169, Japan}
\affiliation{Wayne State University, Detroit, Michigan 48201, USA}
\affiliation{University of Wisconsin, Madison, Wisconsin 53706, USA}
\affiliation{Yale University, New Haven, Connecticut 06520, USA}

\author{T.~Aaltonen}
\affiliation{Division of High Energy Physics, Department of Physics, University of Helsinki and Helsinki Institute of Physics, FIN-00014, Helsinki, Finland}
\author{S.~Amerio}
\affiliation{Istituto Nazionale di Fisica Nucleare, Sezione di Padova-Trento, $^{ff}$University of Padova, I-35131 Padova, Italy}
\author{D.~Amidei}
\affiliation{University of Michigan, Ann Arbor, Michigan 48109, USA}
\author{A.~Anastassov$^x$}
\affiliation{Fermi National Accelerator Laboratory, Batavia, Illinois 60510, USA}
\author{A.~Annovi}
\affiliation{Laboratori Nazionali di Frascati, Istituto Nazionale di Fisica Nucleare, I-00044 Frascati, Italy}
\author{J.~Antos}
\affiliation{Comenius University, 842 48 Bratislava, Slovakia; Institute of Experimental Physics, 040 01 Kosice, Slovakia}
\author{G.~Apollinari}
\affiliation{Fermi National Accelerator Laboratory, Batavia, Illinois 60510, USA}
\author{J.A.~Appel}
\affiliation{Fermi National Accelerator Laboratory, Batavia, Illinois 60510, USA}
\author{T.~Arisawa}
\affiliation{Waseda University, Tokyo 169, Japan}
\author{A.~Artikov}
\affiliation{Joint Institute for Nuclear Research, RU-141980 Dubna, Russia}
% ...

\collaboration{CDF Collaboration}\altaffiliation[With visitors from]{
$^a$University of British Columbia, Vancouver, BC V6T 1Z1, Canada,
$^b$Istituto Nazionale di Fisica Nucleare, Sezione di Cagliari, 09042 Monserrato (Cagliari), Italy,
$^c$University of California Irvine, Irvine, CA 92697, USA,
$^e$Institute of Physics, Academy of Sciences of the Czech Republic, 182~21, Czech Republic,
$^f$CERN, CH-1211 Geneva, Switzerland,
$^g$Cornell University, Ithaca, NY 14853, USA,
$^{dd}$The University of Jordan, Amman 11942, Jordan,
$^h$University of Cyprus, Nicosia CY-1678, Cyprus,
$^i$Office of Science, U.S. Department of Energy, Washington, DC 20585, USA,
$^j$University College Dublin, Dublin 4, Ireland,
$^k$ETH, 8092 Z\"{u}rich, Switzerland,
$^l$University of Fukui, Fukui City, Fukui Prefecture, Japan 910-0017,
$^m$Universidad Iberoamericana, Lomas de Santa Fe, M\'{e}xico, C.P. 01219, Distrito Federal,
$^n$University of Iowa, Iowa City, IA 52242, USA,
$^o$Kinki University, Higashi-Osaka City, Japan 577-8502,
$^p$Kansas State University, Manhattan, KS 66506, USA,
$^q$Brookhaven National Laboratory, Upton, NY 11973, USA,
$^r$University of Manchester, Manchester M13 9PL, United Kingdom,
$^s$Queen Mary, University of London, London, E1 4NS, United Kingdom,
$^t$University of Melbourne, Victoria 3010, Australia,
$^u$Muons, Inc., Batavia, IL 60510, USA,
$^v$Nagasaki Institute of Applied Science, Nagasaki 851-0193, Japan,
$^w$National Research Nuclear University, Moscow 115409, Russia,
$^x$Northwestern University, Evanston, IL 60208, USA,
$^y$University of Notre Dame, Notre Dame, IN 46556, USA,
$^z$Universidad de Oviedo, E-33007 Oviedo, Spain,
$^{aa}$CNRS-IN2P3, Paris, F-75205 France,
$^{cc}$Universidad Tecnica Federico Santa Maria, 110v Valparaiso, Chile,
$^{ll}$Universite catholique de Louvain, 1348 Louvain-La-Neuve, Belgium,
$^{oo}$University of Z\"{u}rich, 8006 Z\"{u}rich, Switzerland,
$^{pp}$Massachusetts General Hospital and Harvard Medical School, Boston, MA 02114 USA,
$^{qq}$Hampton University, Hampton, VA 23668, USA,
$^{rr}$Los Alamos National Laboratory, Los Alamos, NM 87544, USA
}
\noaffiliation



\begin{abstract} 
We present the first signature-based search for delayed photons using an exclusive photon plus missing transverse energy final state. Events are reconstructed
in a data sample from the CDF II detector corresponding to $6.3\ \fb$ of integrated luminosity from $\sqrt{s}=1.96$~TeV proton-antiproton collisions.  
Candidate events are selected if they contain a photon with an arrival time in the detector larger than
expected from a promptly-produced photon.
The mean number of events from standard model sources predicted by the data-driven background model 
based on the photon timing distribution is $286 \pm 24$. A total of 322 events are observed.
A $p$-value of $12\%$ is obtained, showing consistency of the data with standard model predictions.


\end{abstract}



\maketitle

\normalsize

The unknown nature of possible particles or interactions beyond the standard model (SM) motivates search strategies at particle collider experiments that do not rely on features of specific phenomenological models, but rather seek generic deviations from the SM expectations~\cite{GlashowPartialSym,*WeinbergLeptons,*SalamRef}. We report a signature-based search in exclusive photon and missing transverse energy ($\MET$)~\cite{CDFCoord} events from proton-antiproton ($\ppbar$) collisions at $\sqrt{s} = 1.96$~GeV where candidate events are selected based on photons that arrive late in the detector relative to the time expected from prompt production ({\it delayed} photon). This would be the signature of a heavy, neutral, long-lived particle that traverses part of the detector and then decays to a photon and a neutral, non-interacting particle that would appear in the detector as $\MET$~\cite{SearchProspect,GMSBHiggsProspect}. Such particles would exist, for example, in gauge-mediated supersymmetry-breaking scenarios ({\it e.g.}, $\neutralinoOne \to \gamma \gravitino$, where $\neutralinoOne$ is the lightest neutralino and $\gravitino$ is the gravitino)~\cite{Dine1982227,*Nappi1982175,*AlvarezGaume198296,*Dimopoulos1983479}. Searches that focus on particular models at collider experiments, where supersymmetric particles appear at the end of a decay chain, found no evidence for these particles~\cite{incDelayedPRL,*incDelayedPRD,springerlink:10.1007/s10052-002-1005-z,*2652040520070420,*Pasztor:918375,*AbdallahDELPHI,*AaltonenGMSBDiphoton,*AbazovDiphoton,
*PhysRevLett.106.211802,*PhysRevLett.106.121803,*CMSDiphoton}. 
We focus on the exclusive $\gamma+\MET$ final state with delayed photons. Since the sensitivity to such scenarios can vary significantly as a function of the model parameters ({\it e.g.}, production mechanism as well as the mass and lifetime of the heavy, non-SM particle)~\cite{SearchProspect,GMSBHiggsProspect}, we conduct a signature-based search and present the results without any optimization or limit interpretation with respect to
a particular model.

This paper summarizes the first such search 
and uses data from $6.3 \text{ fb}^{-1}$ of integrated
luminosity collected 
with the CDF II detector at the Fermilab Tevatron~\cite{JonThesis,*AdamThesis}. 
An important variable in this search is the difference between the observed arrival time of a photon in the detector
and the time predicted for photons promptly produced in the primary $\ppbar$ interaction. This difference in time, $\deltat$, is used to distinguish signal candidate events from both SM-collision and non-collision background sources.
For photons produced in decays of heavy, long-lived particles, the distribution in $\deltat$ would be shifted towards positive values~\cite{SearchProspect}. 
A full estimation of the backgrounds to the photon and missing energy final state is performed in a data-driven manner, and compared
to data to determine whether any significant excess of events exists.

Detailed descriptions of the CDF II detector can be found elsewhere~\cite{CDFII}.
The detector subsystems relevant to this analysis are briefly mentioned here.  The event kinematic properties and detector geometry are described in a cylindrical 
coordinate system~\cite{CDFCoord}. The detector is composed of a silicon microstrip tracking 
system (``silicon vertex detector"), a tracking drift chamber, a calorimeter
detector, and a muon detector. The silicon vertex detector provides a high-precision position measurement 
of charged-particle trajectories~\cite{SVX}, while the drift chamber provides accurate momentum measurements and allows the reconstruction
of each charged particle's production time~\cite{COT}. The combination of these measurements provides accurate reconstruction of the
position ($\vec{x_{\text{i}}}$) and time ($t_\text{i}$) of the primary $\ppbar$ interaction. 
The $\ppbar$ luminous region is approximately described by a Gaussian 
distribution centered at $z_{\text{i}} = 0$ with an rms spread of $28$~cm and mean time of
$t_{\text{i}} = 0$ with an rms spread of $1.28$~ns.
The $\ppbar$ interactions are reconstructed using an algorithm that 
combines well measured tracks to form a candidate 
vertex~\cite{incDelayedPRL,*incDelayedPRD}. Vertex candidates must consist of at
least three high-quality tracks that intersect each other within $1.5$~cm along the $z$-axis and
within $1.5$~ns in time, with $\Sigma p_T \geq 5$~GeV and $\left|z\right| < 60$~cm, 
where $\sum p_T$ is the scalar sum of the transverse momenta of the corresponding charged particles. 
The reconstructed vertices
use the average $z$ and $t$ of the tracks, and
have a spatial resolution of
$0.24$~cm in $z$ and a time resolution of $0.22$~ns.

The calorimeter has a pointing-tower geometry and is composed of separate electromagnetic
and hadronic compartments that are 
used to identify photons, electrons, jets, and muons, as well as measure $\MET$ in the event. 
The measurement of the arrival time ($t_{\text{f}}$) of photons (and electrons) in the electromagnetic calorimeter is done using  a fixed-threshold discriminator and a time-to-digital converter (TDC) system~\cite{EMTiming} which is connected to each electromagnetic tower, and has a resolution of $0.60$~ns. 
The measurement of the arrival position ($\vec{x}_\text{f}$) is measured by the shower-maximum detector in the electromagnetic calorimeter and has a resolution of 0.2~cm~\cite{CDFCalorimeter}.


The CDF experiment uses a multi-trigger online data acquisition system. This analysis uses events
selected with a trigger that requires a photon candidate having
at least $25$~GeV of $\et$, in addition to a requirement
of at least $25$~GeV of $\MET$ in the event. 
By also allowing candidate events from one or more additional photon triggers~\cite{incDelayedPRL,*incDelayedPRD},
we achieve approximately $100\%$ efficiency for events passing the final selection 
requirements~\cite{incDelayedPRL,*incDelayedPRD}.  

In the offline analysis, photon
candidates are required to meet standard photon identification requirements with a minor modification as described in Ref.~\cite{incDelayedPRL,*incDelayedPRD},
to retain efficiency for photons that do not come directly from the beam line. 
For reasons described below, the offline photon $\et$ and event $\MET$ values are calculated with respect to the center of the detector ($\Etzero$ and $\METzero$) rather than the 
selected primary vertex. Backgrounds from non-collision sources (cosmic rays and beam-halo sources) are 
rejected using standard criteria~\cite{incDelayedPRL,*incDelayedPRD,Aaltonen:2008hh} along with 
new requirements~\cite{CosmicRayReqs} that render the beam-halo background negligible.

Collision backgrounds in the exclusive $\gamma+\MET$ data sample result from
processes of $\gamma+\text{jets}$ production, where unreconstructed jets mimic $\MET$; $Z\gamma\to\nu\nu\gamma$ production;
$W\to l\nu$ production, where the lepton or an extraneous jet is misidentified as a photon;
and $W(\gamma)\to l\nu\gamma$ production, where the lepton is not identified.
Raising the $\Etzero$ and $\METzero$ thresholds to 45~GeV and requiring the exclusive $\gamma+\MET$ final state rejects most of these background; 
any event with a reconstructed track with $p_T > 10$~GeV/c and rapidity magnitude $\left|\eta\right|<1.6$~\cite{CDFCoord} is
excluded from the analysis. 
Similarly, events are rejected if they contain an additional energy cluster, reconstructed with the {\sc jetclu} algorithm~\cite{jetID} with a $\Delta R = 0.4$ cone, with $\Et > 15$~GeV.


We calculate $\deltat$ for each photon candidate using:

\begin{equation}\label{eqn:tcorr}
	\deltat = \left(t_{\text{f}} - t_{\text{i}}\right) - \text{TOF},
\end{equation}
where $\text{TOF}=\left(\left| \vec{x}_{\text{f}} - \vec{x}_{\text{i}}\right|\right)/c$ 
is the expected time-of-flight of a prompt photon from the selected $\ppbar$ interaction vertex
to the location of the associated energy deposit in the calorimeter.
For a promptly-produced photon, $\deltat = 0$~ns in a hypothetical detector with perfect timing resolution.
The signal region for this analysis
is defined as $ 2 < \deltat < 7$~ns to remove most promptly-produced photons  
at small values of $\deltat$ and cosmic-ray events at large times, but retain heavy, long-lived particles that would have
decayed before leaving the detector~\cite{SearchProspect}. If multiple vertices are reconstructed in the event, the vertex with the highest $\sum p_T$ is selected as the primary vertex.

The background contributions from non-collision and collision sources are estimated from data.  The non-collision backgrounds are dominated by cosmic-ray sources~\cite{incDelayedPRL,*incDelayedPRD} that are distributed uniformly in time. They are modeled using a data-driven background estimate using events in the region of $20<\deltat<80$~ns and an extrapolation
into the signal region. The collision backgrounds can be divided into two classes of events.  The first class includes events in which the photon is correctly associated with its production vertex and are readily reduced by the final timing requirement. The second class includes events in which the primary vertex is incorrectly selected as the production vertex of the photon. 

The dominant collision background in the signal region comes from 
prompt SM-photon production events in which the photon is associated with the wrong primary vertex. 
A wrong vertex assignment can occur either because the $\ppbar$ interaction that
produced the photon was not reconstructed or because an additional 
$\ppbar$ interaction produced another vertex that was mistakenly associated with the photon. 
Although it can only be done on a statistical basis, each collision-background event can be classified as a {\it right-vertex} or a {\it wrong-vertex} event.
While the probability of an event being a wrong-vertex type is dependent on the number of extra collisions in the event or the instantaneous luminosity, the fraction of each type can be measured in data.

%

Monte Carlo simulations of all the expected SM background processes were performed to aid in their study and rejection.
The simulated samples are
$W\to e\nu_{e}$, $W\to\mu\nu_{\mu}$, $W\to\tau\nu_{\tau}$, 
$\gamma+\text{jet}$, and $Z\gamma\to\nu\nu\gamma$ (all produced
by the {\sc pythia} event generator~\cite{pythia1,*pythia2}, which adds initial- and final-state radiation), 
and $W(\gamma)\to l\nu\gamma$ 
(produced by the {\sc baur} event generator~\cite{baur}).
The detector response in all simulation samples, including multiple collisions in the event, is modeled by a {\sc geant}-based detector simulation~\cite{CDFSimGeant} and allows each event to be classified as a right-vertex or wrong-vertex event.
This analysis also uses a control sample of exclusive $e + \MET$ collision events~\cite{ElectronIDforPRDRC}
because the final state differs from
exclusive $\gamma+\MET$ final state only in the charged-particle
track associated with the electron.  In such events, the
electron track is removed from the event reconstruction to
emulate exclusive $\gamma+\MET$ events, then used {\it a posteriori}
to determine whether the emulated photon is correctly associated with its production vertex.

The $\deltat$ distributions for right-vertex and wrong-vertex events, $\deltatRV$ and $\deltatWV$ respectively,
are both well modeled by Gaussian distributions after all selection requirements.
The distribution describing right-vertex events
has a mean timing of $\langle \deltatRV \rangle = 0.0 \pm 0.05$~ns and
rms spread of $0.65 \pm 0.05$~ns~\cite{EMTiming},  
due to the contributions of the
vertex-reconstruction algorithm and the calorimeter-timing resolutions. 
The distribution describing wrong-vertex collision events has
an rms spread of $2.0\pm 0.1$ ns~\cite{EMTiming}. Its mean depends on the associated
SM processes and cannot be assumed {\it a priori}. 
Since the collision-background timing distribution is described 
by the sum of the right-vertex and wrong-vertex event distributions,
the collective collision background is modeled by the sum of two Gaussian distributions.
Five of the six parameters describing these two Gaussian distributions are directly
determined from the $\gamma+\MET$ candidate-event data sample 
since the region $-2 \lesssim \deltat \lesssim 2$ ns is dominated by right-vertex events, and
the region with large negative times $-7 \lesssim \deltat \lesssim -2$~ns 
is dominated by wrong-vertex events.
An independent measurement, for example $\langle \deltatWV \rangle$, is still needed to correctly model the collision backgrounds.


The crucial element of this analysis is that the wrong-vertex timing distribution is well described by a Gaussian whose mean can be measured in a data-driven manner.
To motivate the additional selection requirements, Eq.~(\ref{eqn:tcorr}) can be rewritten in a form that illustrates the sources of non-Gaussian tails and non-zero mean times for wrong-vertex events. Since $t_\text{f}$ equals $t_{\text{i}}^{\text{R}}+\tofrv$ for SM background events, in the absence of detector effects,
the measured $\deltat$ when a wrong vertex is selected can be approximated as

\begin{equation}\label{eqn:tcorrWV}
	\deltat^{\text{W}} = \left(t_{\text{i}}^{\text{R}} - t_{\text{i}}^{\text{W}}\right) + \left(\tofrv - \tofwv\right),
\end{equation}
where we continue the use of the superscripts R and W for the variables.
The first term in the right-hand side of Eq.~(\ref{eqn:tcorrWV}) has a mean value of zero with an rms spread of $\sqrt{2}\times 1.28 \text{ ns} = 1.8$~ns resulting from the luminous-region parameters. Additional variation due to the measurement of the arrival time brings the full rms spread to 2.0~ns~\cite{EMTiming}. The rms spread of the background-dependent second term is typically smaller than 0.4~ns, but three effects introduce process-specific
non-Gaussian tails and $\mathcal{O}(0.1)$~ns biases on the mean.

The first source of wrong-vertex timing bias is a threshold effect that affects events with the photon $\Et$
near the analysis threshold of $45$~GeV. Use of the wrong vertex position in photon 
reconstruction biases the measured value of the
photon transverse energy ($E_T^\text{m}$) with respect to its 
true value ($E_T^\text{t}$) for geometric reasons.
If $\Et$ is defined as $\Et = E \sin \theta$, where $\theta$ is the photon-momentum polar angle defined with respect to the primary vertex,
selection of the wrong vertex results in misreconstruction of the photon $\Et$.
If the selection of the wrong vertex results in a shorter path length 
from the collision to the calorimeter, then the $\et$ is overestimated 
and the second term in Eq.~(\ref{eqn:tcorrWV}) is positive ($\tofrv>\tofwv$).
Events with photon $E_T^\text{t} < 45$~GeV and $E_T^\text{m} > 45$~GeV  
contribute to the sample with an average $\deltatWV$ value biased
towards larger times.
Similarly, an event with a photon with $E_T^\text{t} > 45$~GeV but 
$E_T^\text{m} < 45$~GeV is removed from the sample because of the negative bias of $\deltatWV$. In this case,
$\deltatWV$ is biased toward negative times. Thus, an $\Et$ threshold biases the $\deltatWV$ distribution towards
positive $\deltatWV$ values.  
In order to minimize this bias, the detector center ($z = 0.0$) is used in computing $E_T$ (and $\MET$ for consistency).


The second source of wrong-vertex timing bias arises because the primary vertex is required to have $\left| z \right| < 60$~cm. Photons in events originating from collisions
which occurred at $\left| z \right| > 60$~cm would necessarily have the wrong vertex used to compute $\deltat$.
This case induces a positive bias in the timing of the second term in the right-hand side of Eq.~(\ref{eqn:tcorrWV}) because the path length from
the selected vertex to the calorimeter is biased to shorter values
than the true path length.
To suppress this source of background, 
the analysis vetoes events with a vertex with $\left| z \right| > 60$~cm using a vertex-identification algorithm that
has high efficiency for collisions at large $\left| z \right|$~\cite{SvxVertexTracker,*SvxVertexTrigger}.
This requirement is $95\%$ efficient for events with a correctly-reconstructed primary
vertex satisfying $\left| z \right| < 60$~cm.

The last significant source of wrong-vertex timing bias is from $W\to e\nu$ events identified as $\gamma+\MET$ in a way that biases $\langle \deltatWV \rangle$
towards positive values. In this case, as the electron traverses the tracking system, it loses most of its energy to a high-energy photon via bremsstrahlung. As the trajectory of the low-energy electron is curved away from the final photon direction, the photon candidate passes all the photon-identification criteria.
In the case that a wrong vertex is selected, the reconstructed photon candidate timing is
biased to a positive value of the second term in the right-hand side of Eq.~(\ref{eqn:tcorrWV}).
This occurs because electrons with a longer path length are more likely to 
generate a photon via bremsstrahlung in the detector, thus the
actual path length traversed is longer on average
than that of a prompt photon produced in the collision.
To reject events in which the electron track does not point to the photon position in the calorimeter, 
an extra requirement is imposed on reconstructed 
tracks that are close to the photon in $\eta-\phi$ space, as measured at the beam line.
An event is rejected if any reconstructed track in it has a value of 
$\sqrt{(\Delta \eta /\sigma_{\eta})^{2} + (\Delta \phi /\sigma_{\phi})^{2}}<5.0$, where 
$\Delta \eta$ and $\Delta \phi$ are the differences between the $\eta$ and $\phi$ of the track and that of the photon, respectively,
and $\sigma_{\eta} = 6.3 \times 10^{-3}$ and 
$\sigma_{\phi} = 8.1 \times 10^{-2}$ are the detector resolutions in
$\Delta \eta$ and $\Delta \phi$, respectively~\cite{JonThesis,*AdamThesis}.
Studies show that this requirement is approximately $95\%$ efficient for prompt photons 
and reduces the background rate from this source by about $70\%$. 

After imposing all of these bias-reducing restructions, the resulting sample contains 5~421 $\gamma+\MET$ 
candidates.

The wrong-vertex events in each MC simulation and $e + \MET$ control sample have a timing distribution that  is 
well modeled by a single Gaussian distribution 
with an rms spread of $2.0\pm 0.1$~ns and a mean that 
varies among the production mechanisms between 0.0 and 0.8 ns. 
With mean variations across samples not exceeding half of the measured rms spread, any combination of the $\deltat$ distributions for the
modeled background processes is found to be well modeled by a single Gaussian distribution with the 
same rms spread within uncertainties. However, the mean value of $\deltatWV$ for the data must be determined separately.

\begin{figure*}
\centering
\includegraphics[width=0.60\linewidth]{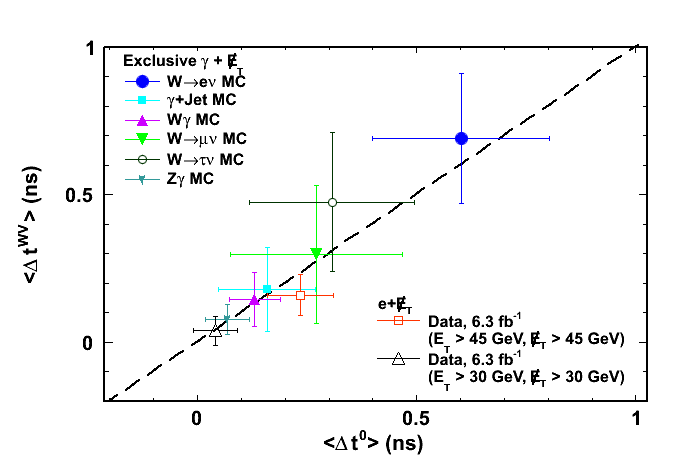}   
\caption{A comparison of $\langle \deltatWV \rangle$ and $\langle \deltatZero \rangle$ in a 
number of Monte Carlo samples that produce the exclusive 
$\gamma+\MET$ final state, and two exclusive $e+\MET$ data samples.}
\label{fig:WV_vs_NV}
\end{figure*}

Since the mean of the wrong-vertex timing events lies in the region dominated by right-vertex events, no fit procedure is sensitive enough to determine $\langle \deltatWV \rangle$ with adequate accuracy.
To determine $\langle \deltatWV \rangle$, a sample of events independent of the exclusive $\gamma + \MET$ sample is created.
This event sample is identical to the $\gamma + \MET$ sample except for a requirement that there be 
no reconstructed vertex, and contains 4~924 events. 
The value of $\deltat$ for an event that does not have a reconstructed
vertex, denoted $\deltatZero$, is computed
assuming an initial time and position of $t_{\text{i}}=0$ and $z_{\text{i}}=0$ in Eq.~(\ref{eqn:tcorr}).
For geometric reasons, $\langle \deltatZero \rangle = \langle \deltatWV \rangle$ 
to a high degree of precision for the entire sample; this is observed
in all simulation and control samples as shown in Fig.~\ref{fig:WV_vs_NV}. The largest discrepancy is 0.08~ns and is taken as the systematic uncertainty on the measurement of $\langle \deltatWV \rangle$.
The $\deltatZero$ distribution is well described by a single Gaussian with an
$\text{rms spread}$ of $1.6 \pm 0.08$~ns and a normalization that is determined from data.

The data from events both with and without a vertex are combined to estimate the full set of backgrounds.
The data outside the signal region is fit to the sum of the two 
Gaussian distributions and the uniform distribution that describe the complete background model.
For the no-vertex sample, a single Gaussian and uniform distribution are used.
A likelihood fit is performed over events with a vertex in the bins spanning
$-7 < \deltat < 2$~ns and $20 < \deltat < 80$~ns,
and for events without a vertex in the bins spanning
$-3.5 < \deltatZero < 3.5$~ns and $20 < \deltatZero < 80$~ns, which are expected to be dominated by collision and cosmic-ray backgrounds.
The likelihood function is defined as a product of Poisson probabilities
over the bins of $\deltat$ and $\deltatZero$ and Gaussian constraints assigned for each systematic
uncertainty.

\begin{figure*}
\centering
\centerline{\subfigure[]{\label{fig:TRaw_StackedFitDeltaTNEW}\includegraphics[width=0.50\linewidth]{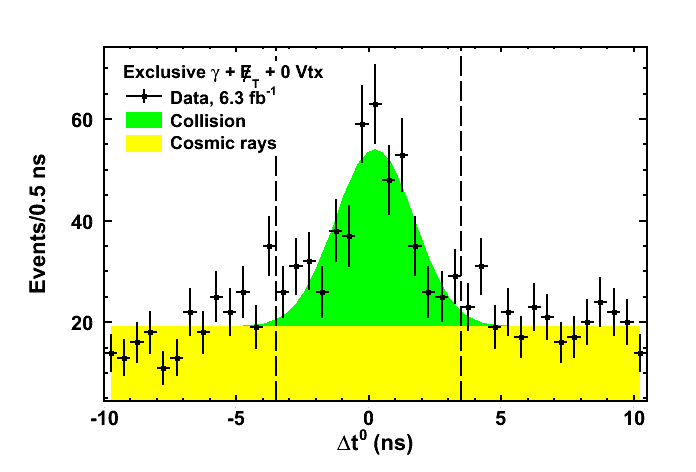}}  
\subfigure[]{\label{fig:TCorr_StackedFit_fullrangeDeltaTNEW}\includegraphics[width=0.50\linewidth]{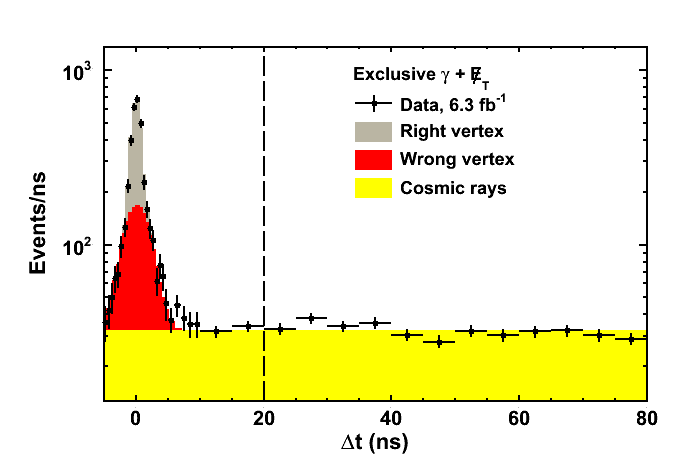}}}
\centerline{\subfigure[]{\label{fig:TCorr_StackedFitDeltaTNEW}\includegraphics[width=0.50\linewidth]{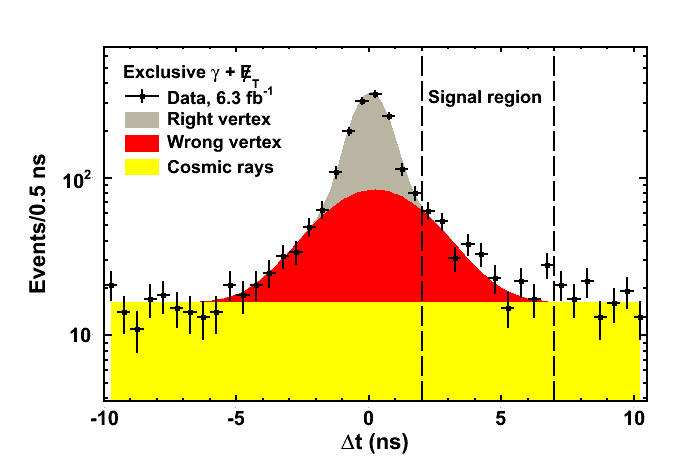}}
\subfigure[]{\label{fig:TCorr_BkgSubDeltaTNEW}\includegraphics[width=0.50\linewidth]{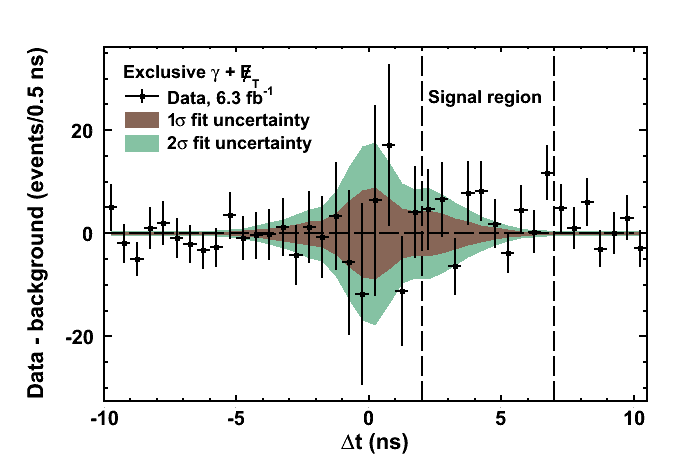}}}
\caption{(a): The $\deltatZero$ timing distribution for $\gamma+\MET$ events with no vertex. The dashed lines demarcate the region $-3.5 < \deltatZero < 3.5$~ns, used to measure $\langle \deltatWV \rangle = \langle \deltatZero \rangle$. (b): The $\deltat$ timing distribution for $\gamma+\MET$ events
with a vertex. The dashed line separates the region $20 < \deltat < 80$~ns, which is used to estimate the rate of cosmic-ray events. (c): The $\deltat$ distribution, from $-10 < \deltat < 10$~ns, with the dashed lines indicating the signal region. In all three cases, the shaded regions reflect the best-fit values determined from the sideband regions. (d): The background-subtracted data distribution with systematic uncertainties.}
\label{fig:novertex}
\end{figure*}

The best-fit values for the two samples (without and with vertices) are shown in 
Figs.~\ref{fig:TRaw_StackedFitDeltaTNEW} and \ref{fig:TCorr_StackedFit_fullrangeDeltaTNEW}. 
A value of $\langle \deltatZero \rangle = \langle \deltatWV \rangle = 0.20\pm 0.13$~ns is obtained from the fit.
Likewise, in the sample of events with a vertex, $875 \pm 66$ right-vertex events, 
$676 \pm 84$ wrong-vertex events, and $31.9 \pm 0.7$~events/ns from cosmic rays are obtained. 
In the signal region, the fit predicts a background of $286\pm 24$ events.
The contributions include $159\pm 4$ cosmic-ray events, $126\pm 24$ 
wrong-vertex events, and $1.0 \pm 0.6$ right-vertex events. 
The uncertainty on the background is dominated by the limited number of events in the sample without a reconstructed vertex. The statistical uncertainty on $\langle \deltatWV \rangle$ produces a 22-event uncertainty on the number of wrong-vertex events in the signal region. The remaining uncertainties are all smaller and are dominated by the systematic uncertainty on the relationship between $\langle \deltatWV \rangle$ and $\langle \deltatZero \rangle$, the uncertainty on the rms spread of $\langle \deltatWV \rangle$, and the uncertainties on the mean and rms spread for $\langle \deltatRV \rangle$.

The $\deltat$ distribution for the data and backgrounds in the region $-10<\deltat<10$~ns is shown in Fig.~\ref{fig:TCorr_StackedFitDeltaTNEW}. The data-minus-background distribution is shown in Fig.~\ref{fig:TCorr_BkgSubDeltaTNEW}. 
A total of $322$ events are observed in the signal region.
The probability for the SM background to yield the observed number of events or more ($p$-value) is determined using simulated experiments that take into account the mean background expectation in the signal region, along with its systematic uncertainty. The resulting $p$-value is 12\%, consistent with standard model-only expectations.


In conclusion, motivated by the possible existence of an unobserved, heavy, long-lived, neutral particle, we present the first signature-based search for the 
production of events with the exclusive photon and 
missing transverse energy final state, where the photon detection time is delayed with respect to the time expected for a photon originated directly from the collision.  We identify a 
number of kinematic properties and detector
effects that can mimic the presence of 
a signal
and use novel analysis techniques to minimize their impact on the results.
We observe no evidence of delayed-photon production in this final state.



\pagebreak
We thank the Fermilab staff and the technical staffs of the participating institutions for their vital contributions. This work was supported by the U.S. Department of Energy and National Science Foundation; the Italian Istituto Nazionale di Fisica Nucleare; the Ministry of Education, Culture, Sports, Science and Technology of Japan; the Natural Sciences and Engineering Research Council of Canada; the National Science Council of the Republic of China; the Swiss National Science Foundation; the A.P. Sloan Foundation; the Bundesministerium f\"ur Bildung und Forschung, Germany; the Korean World Class University Program, the National Research Foundation of Korea; the Science and Technology Facilities Council and the Royal Society, UK; the Russian Foundation for Basic Research; the Ministerio de Ciencia e Innovaci\'{o}n, and Programa Consolider-Ingenio 2010, Spain; the Slovak R\&D Agency; the Academy of Finland; the Australian Research Council (ARC); and the EU community Marie Curie Fellowship contract 302103.


\bibliography{PhotonMetCitations}

\ifx\mcitethebibliography\mciteundefinedmacro
\PackageError{apsrevM.bst}{mciteplus.sty has not been loaded}
{This bibstyle requires the use of the mciteplus package.}\fi
\begin{mcitethebibliography}{38}
\expandafter\ifx\csname natexlab\endcsname\relax\def\natexlab#1{#1}\fi
\expandafter\ifx\csname bibnamefont\endcsname\relax
  \def\bibnamefont#1{#1}\fi
\expandafter\ifx\csname bibfnamefont\endcsname\relax
  \def\bibfnamefont#1{#1}\fi
\expandafter\ifx\csname citenamefont\endcsname\relax
  \def\citenamefont#1{#1}\fi
\expandafter\ifx\csname url\endcsname\relax
  \def\url#1{\texttt{#1}}\fi
\expandafter\ifx\csname urlprefix\endcsname\relax\def\urlprefix{URL }\fi
\providecommand{\bibinfo}[2]{#2}
\providecommand{\eprint}[2][]{\url{#2}}

\bibitem[{\citenamefont{Glashow}(1961)}]{GlashowPartialSym}
\bibinfo{author}{\bibfnamefont{S.~L.} \bibnamefont{Glashow}},
  \bibinfo{journal}{Nucl. Phys.} \textbf{\bibinfo{volume}{22}},
  \bibinfo{pages}{579} (\bibinfo{year}{1961})\relax
\mciteBstWouldAddEndPuncttrue
\mciteSetBstMidEndSepPunct{\mcitedefaultmidpunct}
{\mcitedefaultendpunct}{\mcitedefaultseppunct}\relax
\EndOfBibitem
\bibitem[{\citenamefont{Weinberg}(1967)}]{WeinbergLeptons}
\bibinfo{author}{\bibfnamefont{S.}~\bibnamefont{Weinberg}},
  \bibinfo{journal}{Phys. Rev. Lett.} \textbf{\bibinfo{volume}{19}},
  \bibinfo{pages}{1264} (\bibinfo{year}{1967})\relax
\mciteBstWouldAddEndPuncttrue
\mciteSetBstMidEndSepPunct{\mcitedefaultmidpunct}
{\mcitedefaultendpunct}{\mcitedefaultseppunct}\relax
\EndOfBibitem
\bibitem[{Sal()}]{SalamRef}
\bibinfo{note}{{A. Salam, {\it Proc. 8th Nobel Symposium, Stockholm}
  (1979)}}\relax
\mciteBstWouldAddEndPuncttrue
\mciteSetBstMidEndSepPunct{\mcitedefaultmidpunct}
{\mcitedefaultendpunct}{\mcitedefaultseppunct}\relax
\EndOfBibitem
\bibitem[{CDF({\natexlab{a}})}]{CDFCoord}
\bibinfo{note}{In the CDF II detector, a particle's direction is characterized
  by the azimuthal angle $\phi$ and the pseudorapidity $\eta = -\ln [ \tan
  (\theta/2) ]$, where the polar angle $\theta$ is measured with respect to the
  proton-beam direction. The transverse energy $E_T$ is defined as $E \sin
  \theta$, where $E$ is the energy in the electromagnetic and hadronic
  calorimeter towers associated with a cluster of energy deposition. The
  missing transverse energy $\vec{\MET}$ is defined as $\vec{\MET} = -\Sigma_i
  E_T^i \vec{n}_i$, where $\vec{n}_i$ is a unit vector in the transverse plane
  that points from the interaction vertex to the $i$th calorimeter tower.
  $\MET$ is the magnitude of $\vec{\MET}$. The cylindrical axis of the detector
  is denoted as the $z$ coordinate and increases positively in the direction of
  the proton beam.}\relax
\mciteBstWouldAddEndPunctfalse
\mciteSetBstMidEndSepPunct{\mcitedefaultmidpunct}
{}{\mcitedefaultseppunct}\relax
\EndOfBibitem
\bibitem[{\citenamefont{Toback and Wagner}(2004)}]{SearchProspect}
\bibinfo{author}{\bibfnamefont{D.~A.} \bibnamefont{Toback}} \bibnamefont{and}
  \bibinfo{author}{\bibfnamefont{P.}~\bibnamefont{Wagner}},
  \bibinfo{journal}{Phys. Rev. D} \textbf{\bibinfo{volume}{70}},
  \bibinfo{pages}{114032} (\bibinfo{year}{2004})\relax
\mciteBstWouldAddEndPuncttrue
\mciteSetBstMidEndSepPunct{\mcitedefaultmidpunct}
{\mcitedefaultendpunct}{\mcitedefaultseppunct}\relax
\EndOfBibitem
\bibitem[{\citenamefont{Mason and Toback}(2011)}]{GMSBHiggsProspect}
\bibinfo{author}{\bibfnamefont{J.~D.} \bibnamefont{Mason}} \bibnamefont{and}
  \bibinfo{author}{\bibfnamefont{D.}~\bibnamefont{Toback}},
  \bibinfo{journal}{Phys. Lett. B} \textbf{\bibinfo{volume}{702}},
  \bibinfo{pages}{377} (\bibinfo{year}{2011})\relax
\mciteBstWouldAddEndPuncttrue
\mciteSetBstMidEndSepPunct{\mcitedefaultmidpunct}
{\mcitedefaultendpunct}{\mcitedefaultseppunct}\relax
\EndOfBibitem
\bibitem[{\citenamefont{Dine and Fischler}(1982)}]{Dine1982227}
\bibinfo{author}{\bibfnamefont{M.}~\bibnamefont{Dine}} \bibnamefont{and}
  \bibinfo{author}{\bibfnamefont{W.}~\bibnamefont{Fischler}},
  \bibinfo{journal}{Phys. Lett. B} \textbf{\bibinfo{volume}{110}},
  \bibinfo{pages}{227 } (\bibinfo{year}{1982})\relax
\mciteBstWouldAddEndPuncttrue
\mciteSetBstMidEndSepPunct{\mcitedefaultmidpunct}
{\mcitedefaultendpunct}{\mcitedefaultseppunct}\relax
\EndOfBibitem
\bibitem[{\citenamefont{Nappi and Ovrut}(1982)}]{Nappi1982175}
\bibinfo{author}{\bibfnamefont{C.~R.} \bibnamefont{Nappi}} \bibnamefont{and}
  \bibinfo{author}{\bibfnamefont{B.~A.} \bibnamefont{Ovrut}},
  \bibinfo{journal}{Phys. Lett. B} \textbf{\bibinfo{volume}{113}},
  \bibinfo{pages}{175 } (\bibinfo{year}{1982})\relax
\mciteBstWouldAddEndPuncttrue
\mciteSetBstMidEndSepPunct{\mcitedefaultmidpunct}
{\mcitedefaultendpunct}{\mcitedefaultseppunct}\relax
\EndOfBibitem
\bibitem[{\citenamefont{Alvarez-Gaum\'e
  et~al.}(1982)\citenamefont{Alvarez-Gaum\'e, Claudson, and
  Wise}}]{AlvarezGaume198296}
\bibinfo{author}{\bibfnamefont{L.}~\bibnamefont{Alvarez-Gaum\'e}},
  \bibinfo{author}{\bibfnamefont{M.}~\bibnamefont{Claudson}}, \bibnamefont{and}
  \bibinfo{author}{\bibfnamefont{M.~B.} \bibnamefont{Wise}},
  \bibinfo{journal}{Nucl. Phys.} \textbf{\bibinfo{volume}{B207}},
  \bibinfo{pages}{96 } (\bibinfo{year}{1982})\relax
\mciteBstWouldAddEndPuncttrue
\mciteSetBstMidEndSepPunct{\mcitedefaultmidpunct}
{\mcitedefaultendpunct}{\mcitedefaultseppunct}\relax
\EndOfBibitem
\bibitem[{\citenamefont{Dimopoulos and Raby}(1983)}]{Dimopoulos1983479}
\bibinfo{author}{\bibfnamefont{S.}~\bibnamefont{Dimopoulos}} \bibnamefont{and}
  \bibinfo{author}{\bibfnamefont{S.}~\bibnamefont{Raby}},
  \bibinfo{journal}{Nucl. Phys.} \textbf{\bibinfo{volume}{B219}},
  \bibinfo{pages}{479 } (\bibinfo{year}{1983})\relax
\mciteBstWouldAddEndPuncttrue
\mciteSetBstMidEndSepPunct{\mcitedefaultmidpunct}
{\mcitedefaultendpunct}{\mcitedefaultseppunct}\relax
\EndOfBibitem
\bibitem[{\citenamefont{{A. Abulencia {\it et
  al}.}}(2007{\natexlab{a}})}]{incDelayedPRL}
\bibinfo{author}{\bibnamefont{{A. Abulencia {\it et al}.}}}
  (\bibinfo{collaboration}{CDF Collaboration}), \bibinfo{journal}{Phys. Rev.
  Lett.} \textbf{\bibinfo{volume}{99}}, \bibinfo{pages}{121801}
  (\bibinfo{year}{2007}{\natexlab{a}})\relax
\mciteBstWouldAddEndPuncttrue
\mciteSetBstMidEndSepPunct{\mcitedefaultmidpunct}
{\mcitedefaultendpunct}{\mcitedefaultseppunct}\relax
\EndOfBibitem
\bibitem[{\citenamefont{{T. Aaltonen {\it et
  al}.}}(2008{\natexlab{a}})}]{incDelayedPRD}
\bibinfo{author}{\bibnamefont{{T. Aaltonen {\it et al}.}}}
  (\bibinfo{collaboration}{CDF Collaboration}), \bibinfo{journal}{Phys. Rev. D}
  \textbf{\bibinfo{volume}{78}}, \bibinfo{pages}{032015}
  (\bibinfo{year}{2008}{\natexlab{a}})\relax
\mciteBstWouldAddEndPuncttrue
\mciteSetBstMidEndSepPunct{\mcitedefaultmidpunct}
{\mcitedefaultendpunct}{\mcitedefaultseppunct}\relax
\EndOfBibitem
\bibitem[{\citenamefont{{A. Heister {\it et
  al}.}}(2002)}]{springerlink:10.1007/s10052-002-1005-z}
\bibinfo{author}{\bibnamefont{{A. Heister {\it et al}.}}}
  (\bibinfo{collaboration}{ALEPH Collaboration}), \bibinfo{journal}{Eur. Phys.
  J. C} \textbf{\bibinfo{volume}{25}}, \bibinfo{pages}{339}
  (\bibinfo{year}{2002})\relax
\mciteBstWouldAddEndPuncttrue
\mciteSetBstMidEndSepPunct{\mcitedefaultmidpunct}
{\mcitedefaultendpunct}{\mcitedefaultseppunct}\relax
\EndOfBibitem
\bibitem[{\citenamefont{Gataullin et~al.}(2007)\citenamefont{Gataullin, Rosier,
  Xia, and Yang}}]{2652040520070420}
\bibinfo{author}{\bibfnamefont{M.}~\bibnamefont{Gataullin}},
  \bibinfo{author}{\bibfnamefont{S.}~\bibnamefont{Rosier}},
  \bibinfo{author}{\bibfnamefont{L.}~\bibnamefont{Xia}}, \bibnamefont{and}
  \bibinfo{author}{\bibfnamefont{H.}~\bibnamefont{Yang}}, \bibinfo{journal}{AIP
  Conference Proceedings} \textbf{\bibinfo{volume}{903}}, \bibinfo{pages}{217 }
  (\bibinfo{year}{2007})\relax
\mciteBstWouldAddEndPuncttrue
\mciteSetBstMidEndSepPunct{\mcitedefaultmidpunct}
{\mcitedefaultendpunct}{\mcitedefaultseppunct}\relax
\EndOfBibitem
\bibitem[{Pas()}]{Pasztor:918375}
\bibinfo{note}{{G.~P\'asztor, PoS HEP2005 346 (2005)}}\relax
\mciteBstWouldAddEndPuncttrue
\mciteSetBstMidEndSepPunct{\mcitedefaultmidpunct}
{\mcitedefaultendpunct}{\mcitedefaultseppunct}\relax
\EndOfBibitem
\bibitem[{\citenamefont{{J. Abdallah {\it et al}.}}(2009)}]{AbdallahDELPHI}
\bibinfo{author}{\bibnamefont{{J. Abdallah {\it et al}.}}}
  (\bibinfo{collaboration}{DELPHI Collaboration}), \bibinfo{journal}{Eur. Phys.
  J. C} \textbf{\bibinfo{volume}{60}}, \bibinfo{pages}{17}
  (\bibinfo{year}{2009})\relax
\mciteBstWouldAddEndPuncttrue
\mciteSetBstMidEndSepPunct{\mcitedefaultmidpunct}
{\mcitedefaultendpunct}{\mcitedefaultseppunct}\relax
\EndOfBibitem
\bibitem[{\citenamefont{{T. Aaltonen {\it et
  al}.}}(2010)}]{AaltonenGMSBDiphoton}
\bibinfo{author}{\bibnamefont{{T. Aaltonen {\it et al}.}}}
  (\bibinfo{collaboration}{CDF Collaboration}), \bibinfo{journal}{Phys. Rev.
  Lett.} \textbf{\bibinfo{volume}{104}}, \bibinfo{pages}{011801}
  (\bibinfo{year}{2010})\relax
\mciteBstWouldAddEndPuncttrue
\mciteSetBstMidEndSepPunct{\mcitedefaultmidpunct}
{\mcitedefaultendpunct}{\mcitedefaultseppunct}\relax
\EndOfBibitem
\bibitem[{\citenamefont{{V.~M. Abazov {\it et al}.}}(2010)}]{AbazovDiphoton}
\bibinfo{author}{\bibnamefont{{V.~M. Abazov {\it et al}.}}}
  (\bibinfo{collaboration}{D0 Collaboration}), \bibinfo{journal}{Phys. Rev.
  Lett.} \textbf{\bibinfo{volume}{105}}, \bibinfo{pages}{221802}
  (\bibinfo{year}{2010})\relax
\mciteBstWouldAddEndPuncttrue
\mciteSetBstMidEndSepPunct{\mcitedefaultmidpunct}
{\mcitedefaultendpunct}{\mcitedefaultseppunct}\relax
\EndOfBibitem
\bibitem[{\citenamefont{{S. Chatrchyan {\it et
  al}.}}(2011)}]{PhysRevLett.106.211802}
\bibinfo{author}{\bibnamefont{{S. Chatrchyan {\it et al}.}}}
  (\bibinfo{collaboration}{CMS Collaboration}), \bibinfo{journal}{Phys. Rev.
  Lett.} \textbf{\bibinfo{volume}{106}}, \bibinfo{pages}{211802}
  (\bibinfo{year}{2011})\relax
\mciteBstWouldAddEndPuncttrue
\mciteSetBstMidEndSepPunct{\mcitedefaultmidpunct}
{\mcitedefaultendpunct}{\mcitedefaultseppunct}\relax
\EndOfBibitem
\bibitem[{\citenamefont{{G. Aad {\it et al}. }}(2011)}]{PhysRevLett.106.121803}
\bibinfo{author}{\bibnamefont{{G. Aad {\it et al}. }}}
  (\bibinfo{collaboration}{ATLAS Collaboration}), \bibinfo{journal}{Phys. Rev.
  Lett.} \textbf{\bibinfo{volume}{106}}, \bibinfo{pages}{121803}
  (\bibinfo{year}{2011})\relax
\mciteBstWouldAddEndPuncttrue
\mciteSetBstMidEndSepPunct{\mcitedefaultmidpunct}
{\mcitedefaultendpunct}{\mcitedefaultseppunct}\relax
\EndOfBibitem
\bibitem[{\citenamefont{{S. Chatrchyan {\it et al}.}}(2013)}]{CMSDiphoton}
\bibinfo{author}{\bibnamefont{{S. Chatrchyan {\it et al}.}}}
  (\bibinfo{collaboration}{CMS Collaboration}), \bibinfo{journal}{Phys. Lett.
  B} \textbf{\bibinfo{volume}{722}}, \bibinfo{pages}{273}
  (\bibinfo{year}{2013})\relax
\mciteBstWouldAddEndPuncttrue
\mciteSetBstMidEndSepPunct{\mcitedefaultmidpunct}
{\mcitedefaultendpunct}{\mcitedefaultseppunct}\relax
\EndOfBibitem
\bibitem[{Jon()}]{JonThesis}
\bibinfo{note}{{J. Asaadi, Ph.D. thesis, Texas A\&M University
  [FERMILAB-THESIS-2012-43]}}\relax
\mciteBstWouldAddEndPuncttrue
\mciteSetBstMidEndSepPunct{\mcitedefaultmidpunct}
{\mcitedefaultendpunct}{\mcitedefaultseppunct}\relax
\EndOfBibitem
\bibitem[{Ada()}]{AdamThesis}
\bibinfo{note}{{A. Aurisano, Ph.D. thesis, Texas A\&M University
  [FERMILAB-THESIS-2012-35]}}\relax
\mciteBstWouldAddEndPuncttrue
\mciteSetBstMidEndSepPunct{\mcitedefaultmidpunct}
{\mcitedefaultendpunct}{\mcitedefaultseppunct}\relax
\EndOfBibitem
\bibitem[{\citenamefont{{A. Abulencia {\it et
  al}.}}(2007{\natexlab{b}})}]{CDFII}
\bibinfo{author}{\bibnamefont{{A. Abulencia {\it et al}.}}}
  (\bibinfo{collaboration}{CDF Collaboration}), \bibinfo{journal}{J. Phys. G}
  \textbf{\bibinfo{volume}{34}}, \bibinfo{pages}{2457}
  (\bibinfo{year}{2007}{\natexlab{b}})\relax
\mciteBstWouldAddEndPuncttrue
\mciteSetBstMidEndSepPunct{\mcitedefaultmidpunct}
{\mcitedefaultendpunct}{\mcitedefaultseppunct}\relax
\EndOfBibitem
\bibitem[{\citenamefont{{A. Sill {\it et al}.}}(2000)}]{SVX}
\bibinfo{author}{\bibnamefont{{A. Sill {\it et al}.}}}
  (\bibinfo{collaboration}{CDF Collaboration}), \bibinfo{journal}{Nucl.
  Instrum. Methods A} \textbf{\bibinfo{volume}{447}}, \bibinfo{pages}{1}
  (\bibinfo{year}{2000})\relax
\mciteBstWouldAddEndPuncttrue
\mciteSetBstMidEndSepPunct{\mcitedefaultmidpunct}
{\mcitedefaultendpunct}{\mcitedefaultseppunct}\relax
\EndOfBibitem
\bibitem[{\citenamefont{{T. Affolder {\it et al}.}}(2004)}]{COT}
\bibinfo{author}{\bibnamefont{{T. Affolder {\it et al}.}}}
  (\bibinfo{collaboration}{CDF Collaboration}), \bibinfo{journal}{Nucl.
  Instrum. Methods A} \textbf{\bibinfo{volume}{526}}, \bibinfo{pages}{249}
  (\bibinfo{year}{2004})\relax
\mciteBstWouldAddEndPuncttrue
\mciteSetBstMidEndSepPunct{\mcitedefaultmidpunct}
{\mcitedefaultendpunct}{\mcitedefaultseppunct}\relax
\EndOfBibitem
\bibitem[{\citenamefont{{M. Goncharov {\it et al}.}}(2006)}]{EMTiming}
\bibinfo{author}{\bibnamefont{{M. Goncharov {\it et al}.}}},
  \bibinfo{journal}{Nucl. Instrum. Methods A} \textbf{\bibinfo{volume}{565}},
  \bibinfo{pages}{543} (\bibinfo{year}{2006})\relax
\mciteBstWouldAddEndPuncttrue
\mciteSetBstMidEndSepPunct{\mcitedefaultmidpunct}
{\mcitedefaultendpunct}{\mcitedefaultseppunct}\relax
\EndOfBibitem
\bibitem[{\citenamefont{{L. Balka {\it et al}.}}(1988)}]{CDFCalorimeter}
\bibinfo{author}{\bibnamefont{{L. Balka {\it et al}.}}},
  \bibinfo{journal}{Nucl. Instrum. Methods A} \textbf{\bibinfo{volume}{267}},
  \bibinfo{pages}{272} (\bibinfo{year}{1988})\relax
\mciteBstWouldAddEndPuncttrue
\mciteSetBstMidEndSepPunct{\mcitedefaultmidpunct}
{\mcitedefaultendpunct}{\mcitedefaultseppunct}\relax
\EndOfBibitem
\bibitem[{\citenamefont{{T. Aaltonen {\it et
  al}.}}(2008{\natexlab{b}})}]{Aaltonen:2008hh}
\bibinfo{author}{\bibnamefont{{T. Aaltonen {\it et al}.}}}
  (\bibinfo{collaboration}{CDF Collaboration}), \bibinfo{journal}{Phys. Rev.
  Lett.} \textbf{\bibinfo{volume}{101}}, \bibinfo{pages}{181602}
  (\bibinfo{year}{2008}{\natexlab{b}})\relax
\mciteBstWouldAddEndPuncttrue
\mciteSetBstMidEndSepPunct{\mcitedefaultmidpunct}
{\mcitedefaultendpunct}{\mcitedefaultseppunct}\relax
\EndOfBibitem
\bibitem[{Cos()}]{CosmicRayReqs}
\bibinfo{note}{{ Since photons from collisions deposit greater quantities of
  energy in the hadronic calorimeter than do photon candidates from cosmic-ray
  events, we require the energy in the hadronic calorimeter element (HadE)
  immediately behind the electromagnetic calorimeter deposit associated with
  the photon to satisfy HadE $> -0.30 + 0.008 E_{T}^0$. Since photon candidates
  from cosmic-ray sources also deposit energy within the electromagnetic
  calorimeter differently than photons from collisions, we require the energy
  measured in the shower-maximum detector of the electromagnetic calorimeter to
  exceed 20\% of the total energy measured in the electromagnetic calorimeter
  [F. Abe {\it et al}. (CDF Collaboration), Phys. Rev. D {\bf 52}, 4784
  (1995)].}}\relax
\mciteBstWouldAddEndPunctfalse
\mciteSetBstMidEndSepPunct{\mcitedefaultmidpunct}
{}{\mcitedefaultseppunct}\relax
\EndOfBibitem
\bibitem[{\citenamefont{{F. Abe {\it et al.}}}(1992)}]{jetID}
\bibinfo{author}{\bibnamefont{{F. Abe {\it et al.}}}}
  (\bibinfo{collaboration}{CDF Collaboration}), \bibinfo{journal}{Phys. Rev. D}
  \textbf{\bibinfo{volume}{45}}, \bibinfo{pages}{1448}
  (\bibinfo{year}{1992})\relax
\mciteBstWouldAddEndPuncttrue
\mciteSetBstMidEndSepPunct{\mcitedefaultmidpunct}
{\mcitedefaultendpunct}{\mcitedefaultseppunct}\relax
\EndOfBibitem
\bibitem[{\citenamefont{{T. Sj\"{o}strand, P. Ed\'en, C. Friberg, L.
  L\"{o}nnblad, G. Miu, S. Mrenna, and E. Norrbin}}(2001)}]{pythia1}
\bibinfo{author}{\bibnamefont{{T. Sj\"{o}strand, P. Ed\'en, C. Friberg, L.
  L\"{o}nnblad, G. Miu, S. Mrenna, and E. Norrbin}}}, \bibinfo{journal}{Comput.
  Phys. Commun.} \textbf{\bibinfo{volume}{135}}, \bibinfo{pages}{238}
  (\bibinfo{year}{2001})\relax
\mciteBstWouldAddEndPuncttrue
\mciteSetBstMidEndSepPunct{\mcitedefaultmidpunct}
{\mcitedefaultendpunct}{\mcitedefaultseppunct}\relax
\EndOfBibitem
\bibitem[{pyt()}]{pythia2}
\bibinfo{note}{{T. Sj\"{o}strand, S. Mrenna, and P. Skands, J. High Energy
  Phys. 05 (2006) 026}}\relax
\mciteBstWouldAddEndPuncttrue
\mciteSetBstMidEndSepPunct{\mcitedefaultmidpunct}
{\mcitedefaultendpunct}{\mcitedefaultseppunct}\relax
\EndOfBibitem
\bibitem[{\citenamefont{Baur et~al.}(1998)\citenamefont{Baur, Han, and
  Ohnemus}}]{baur}
\bibinfo{author}{\bibfnamefont{U.}~\bibnamefont{Baur}},
  \bibinfo{author}{\bibfnamefont{T.}~\bibnamefont{Han}}, \bibnamefont{and}
  \bibinfo{author}{\bibfnamefont{J.}~\bibnamefont{Ohnemus}},
  \bibinfo{journal}{Phys. Rev. D} \textbf{\bibinfo{volume}{57}},
  \bibinfo{pages}{2823} (\bibinfo{year}{1998})\relax
\mciteBstWouldAddEndPuncttrue
\mciteSetBstMidEndSepPunct{\mcitedefaultmidpunct}
{\mcitedefaultendpunct}{\mcitedefaultseppunct}\relax
\EndOfBibitem
\bibitem[{CDF({\natexlab{b}})}]{CDFSimGeant}
\bibinfo{note}{{R.~Brun, F.~Bruyant, M.~Maire, A.~McPherson, and P~Zanarini,
  CERN Report No. CERN-DD/EE/84-1 (1987); we use the standard CDF simulation
  software as described in E.~Gerchtein and M.~Paulini (2003), arXiv:0306031
  and add a parametrized timing simulation.}}\relax
\mciteBstWouldAddEndPunctfalse
\mciteSetBstMidEndSepPunct{\mcitedefaultmidpunct}
{}{\mcitedefaultseppunct}\relax
\EndOfBibitem
\bibitem[{Ele()}]{ElectronIDforPRDRC}
\bibinfo{note}{{Each electron is required to pass the standard electron
  requirements described in T. Aaltonen {\it et al}. (CDF Collaboration), Phys.
  Rev. D {\bf 77}, 112001 (2008), with the exception of the
  $\chi^{2}_{\text{CES}}$, $\left|\Delta x\right|$ and $\left|\Delta z\right|$
  requirements to make them more consistent with the photon identification
  requirements.}}\relax
\mciteBstWouldAddEndPunctfalse
\mciteSetBstMidEndSepPunct{\mcitedefaultmidpunct}
{}{\mcitedefaultseppunct}\relax
\EndOfBibitem
\bibitem[{\citenamefont{{W. Ashmanskas {\it et al}.}}(2002)}]{SvxVertexTracker}
\bibinfo{author}{\bibnamefont{{W. Ashmanskas {\it et al}.}}},
  \bibinfo{journal}{Nucl. Instrum. Methods A} \textbf{\bibinfo{volume}{477}},
  \bibinfo{pages}{451} (\bibinfo{year}{2002})\relax
\mciteBstWouldAddEndPuncttrue
\mciteSetBstMidEndSepPunct{\mcitedefaultmidpunct}
{\mcitedefaultendpunct}{\mcitedefaultseppunct}\relax
\EndOfBibitem
\bibitem[{\citenamefont{{B. Ashmanskas {\it et al}.}}(2004)}]{SvxVertexTrigger}
\bibinfo{author}{\bibnamefont{{B. Ashmanskas {\it et al}.}}},
  \bibinfo{journal}{Nucl. Instrum. Methods A} \textbf{\bibinfo{volume}{518}},
  \bibinfo{pages}{532} (\bibinfo{year}{2004})\relax
\mciteBstWouldAddEndPuncttrue
\mciteSetBstMidEndSepPunct{\mcitedefaultmidpunct}
{\mcitedefaultendpunct}{\mcitedefaultseppunct}\relax
\EndOfBibitem
\end{mcitethebibliography}

\end{document}